\documentclass[11pt]{article}
\usepackage[utf8]{inputenc}
\usepackage[T1]{fontenc}
\usepackage{tikz}
\usepackage{url}
\usepackage{amsmath}
\usepackage{times}
\usepackage{footmisc}
\usepackage[super]{natbib}

\title{Modeling, state estimation, and optimal control for the US COVID-19 outbreak}

\author
{Calvin Tsay,$^{1\dagger}$ Fernando Lejarza,$^{1\dagger}$ Mark A. Stadtherr $^{1}$, Michael Baldea $^{1,2\ast}$ \\
	\\
	\normalsize{$^{1}$McKetta Department of Chemical Engineering, The University of Texas at Austin,}\\
	\normalsize{Austin, TX, USA}\\
	\normalsize{$^{2}$Oden Institute for Computational Engineering and Sciences, The University of Texas at Austin}\\
	\normalsize{Austin, TX, USA}\\
	\\
	\normalsize{$^\dagger$These authors contributed equally to this work} \\ 
	\normalsize{$^\ast$To whom correspondence should be addressed; E-mail:  mbaldea@che.utexas.edu}
}

\begin{document}

\maketitle

\begin{abstract}
The novel coronavirus SARS-CoV-2 and resulting COVID-19 disease have had an unprecedented spread and continue to cause an increasing number of fatalities worldwide. While vaccines are still under development, social distancing, extensive testing, and quarantining of confirmed infected subjects remain the most effective measures to contain the pandemic.  These measures carry a significant socioeconomic cost. In this work, we introduce a novel optimization-based decision-making framework for managing the COVID-19 outbreak in the US. This includes modeling the dynamics of affected populations, estimating the model parameters and hidden states from data, and an optimal control strategy for sequencing social distancing and testing events such that the number of infections is minimized. The analysis of our extensive computational efforts reveals that social distancing and quarantining are most effective when implemented early, with quarantining of confirmed infected subjects having a much higher impact. Further, we find that ``on-off'' policies alternating between strict social distancing and relaxing such restrictions can be effective at ``flattening'' the curve while likely minimizing social and economic cost. 
\end{abstract}

\section*{Introduction}

Since its first reported case in early December 2019 in Wuhan, Hubei Province, China, the novel Severe Acute Respiratory Syndrome Coronavirus 2 (SARS-CoV-2) and the resulting  COVID-19 disease have reached 184 countries/regions causing a total of 149,024 deaths, as of this writing. The rate of spread of the virus is substantially higher than that of similar previously reported epidemics such as SARS and MERS.\cite{peeri2020sars} Because of this world-wide and rapid spread, on March 11, 2020, the World Health Organization (WHO) declared the COVID-19 outbreak a global pandemic.\cite{sohrabi2020world} The exponential increase in the number of confirmed infectious cases and deaths has caused countries all around the world to respond with severe lock-down, quarantining, and social distancing measures to contain the spread of the disease. For example, northern Italy went into a state of emergency on March 8, 2020, imposing a complete lock-down that was expanded to the rest of the country three days later.\cite{paterlini2020lockdown} Following Italy's attempts to decrease the contagion rate, Spain enforced a nationwide lock-down on March 16. \cite{tobias2020evaluation} In contrast, the response in the US was significantly slower than those of the aforementioned (and most other European) countries, delaying widespread social distancing measures and disease screening initiatives.

Given the exponential growth of virus infections, policy makers face the urgent challenge of determining the appropriate response(s).
For example, restrictions on population mobility, increased resource investment (e.g., personal protective equipment, ventilators, hospital beds in intensive care units), as well as improved COVID-19 screening can greatly impact the spread of infection.\cite{kraemer2020effect, gostic2020estimated}
Here, fundamental epidemiological models,\cite{keeling2011modeling} typically comprising sets of coupled, nonlinear ordinary differential equations (ODEs), are valuable tools for simulating the dynamics of the epidemic and investigating suppression strategies.
Several works (e.g., Anastassopoulou et al., \cite{anastassopoulou2020data} Peng et al., \cite{peng2020} Magal and Webb \cite{ magal2020}) have addressed the development and fitting of such models for the COVID-19 outbreak using regression techniques.
Models were subsequently used to characterize the infection,\cite{kucharski2020early, liu2020reproductive} predict its potential spread,\cite{liu2020predicting, park2020time, wu2020} and/or evaluate mitigation strategies.\cite{sameni2020, pan2020effectiveness}

On the other hand, fewer studies have investigated the dynamic optimization, or \textit{optimal control}, based on epidemiological models in relation to the current COVID-19 outbreak. In this context, the aforementioned epidemiological models are used to determine the policies (i.e., model inputs), such as the social distancing measures and testing rates, that lead to the best outcomes at the population level. Such outcomes include, for example, minimizing the peak number of infected people or total number of deaths, while also accounting for constrained resource availability and the extent to which social distancing is feasible. Among works of this type, Djidjou-Demasse et al. \cite{djidjou2020optimal} investigated the optimal control of a single ``intervention'' input variable for an ODE model.
Moore and Okyere \cite{moore2020controlling} formulated a similar problem with additional inputs, including hospitalization rates and environmental spraying.
Both studies solved the optimal control problem using an iterative forward-backward sweep method \cite{lenhart2007optimal} and assumed the model parameters to be fully known.
While optimal control of epidemiological models has been well-studied (e.g., Biswas et al.,\cite{biswas2014seir} Neilan and Lenhart \cite{ neilan2010introduction}), a limiting factor for implementing modeling and optimization concepts during early stages of an outbreak is that obtaining accurate estimates of key model parameters can be challenging. 

In this work, we report a novel and complete dynamic optimization-based approach to the entire epidemiological modeling and outbreak control workflow for the US COVID-19 outbreak.
We first formulate a dynamic optimization strategy for identifying both the values of time-invariant parameters and the historical trajectories of time-varying parameters (i.e., inputs) of an epidemiological model.
We then investigate how optimal control of the inputs of the same model, which reflect social distancing and testing, relates to infection mitigation strategies.
The optimal control problem is cast as a \textit{simultaneous} dynamic optimization problem (i.e., reformulated as an algebraic problem), \cite{biegler2007overview} enabling the natural use of deterministic global optimization technology.\cite{floudas2013deterministic}
This in turn provides solutions that are proven to be globally optimal, whereas iterative schemes such as the above \cite{lenhart2007optimal} often only guarantee global optimality under certain conditions (the term ``global'' here refers to the notion that the best possible disease control
policy is identified from a set of ``local'' solutions that otherwise satisfy optimality conditions. ``Global'' in this context should not be interpreted as ``world-wide''.
Furthermore, we show how state estimation should be used to update important hidden model states (e.g., the number unconfirmed infections) to reduce the impact of model inaccuracy.

\newpage
Implementations of the parameter estimation and optimal control problems in open-source software are provided freely at \url{https://github.com/Baldea-Group/covid-19}.


\section*{Results}
\subsection*{Mathematical Modeling}
We consider the compartmental model shown in Fig \ref{fig:model}, which brings a few key modifications to the conventional SEIR (susceptible-exposed-infectious-recovered) structure.
Importantly, the $a(t)$ state is added to account for infected subjects that are not included in the count of confirmed cases, either because they are asymptomatic or because of insufficient testing.
The virus is thought to be asymptomatic in 20--40\% of cases \cite{mizumoto2020, nishiura2020} and may be transmitted by asymptomatic carriers. \cite{bai2020}
We also include a $p(t)$ state, which tracks the population that perishes due to the virus.

The modified SEAIR model has six compartments/states: $s(t)$ represents the number of subjects that are susceptible to infection, $e(t)$ the number that have been exposed to the virus, $a(t)$ the number that are infected but asymptomatic/unconfirmed, $i(t)$ the number with confirmed infections, and $r(t)$ the number that have recovered from infection. The states are assumed to sum to a known total population $N$, and $p(t)$ can be calculated algebraically at all times.
The model structure is shown in Fig \ref{fig:model}.

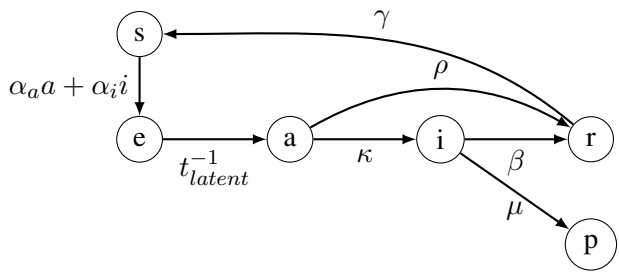
\tikzstyle{arrow} = [thick,->,>=latex]
\tikzstyle{compartment} = [circle, minimum width = 0.6cm, text centered, draw = black, font=\normalsize]
\begin{figure}[!ht]
\centering
\begin{tikzpicture}[node distance = 2cm]
\node(s) [compartment] {s};
\node(e) [compartment, below of=s, yshift = 0.6cm] {e};
\node(a) [compartment, right of=e] {a};
\node(i) [compartment, right of=a] {i};
\node(r) [compartment, right of = i] {r};
\node(p) [compartment, below of=r, yshift = 0.6cm] {p};
\draw [arrow] (s) -- node[anchor=east] {$\alpha_a a + \alpha_i i$} (e);
\draw [arrow] (e) -- node[anchor=north] {$t_{latent}^{-1}$} (a);
\draw [arrow] (a) -- node[anchor=north] {$\kappa$} (i);
\draw [arrow] (i) -- node[anchor=north] {$\beta$} (r);
\draw [arrow] (i) -- node[anchor=north] {$\mu$} (p);
\draw [arrow] (a) to [out = 30, in = 150] node[anchor=south] {$\rho$} (r);
\draw [arrow] (r) to [out = 140, in = 0] node[anchor=south] {$\gamma$} (s);
\end{tikzpicture}
\vspace{10pt}

\begin{tabular}{|l|l|l|}
\hline
{\bf Input/parameter} & {\bf Range} & {\bf Description} \\ \hline
$\alpha_a(t)$ [days$^{-1}$] & $[0.05, 0.5]$ & social distancing \\ \hline
$\alpha_i(t)$ [days$^{-1}$] &  $[0.01, 0.3]$ & quarantining \\ \hline
$\kappa(t)$ [days$^{-1}$] &  $[0.10, 0.3]$ & rate of testing\\ \hline
$\beta$ [days$^{-1}$] & $[0.001, 0.05]$ & recovery rate \\ \hline
$\mu$ [days$^{-1}$] & $[0.001, 0.05 ]$ & death rate\\ \hline
$e_0$ [people] & $[0, N\times10^{-6} ] $ & initial exposed \\  \hline
\end{tabular}
\caption{Digraph representation of SEAIR model; model inputs and parameters.}
\label{fig:model}
\end{figure}

The SEAIR model is described by the following equations:
\begin{align} 
\frac{ds(t)}{dt} &= \frac{-\alpha_a(t)}{N} s(t)a(t) - \frac{\alpha_i(t)}{N} s(t) i(t) + \gamma r(t) \label{eq:ode1} \\
\frac{de(t)}{dt} &= \frac{\alpha_a(t)}{N} s(t)a(t) + \frac{\alpha_i(t)}{N} s(t) i(t) - t_{\text{latent}}^{-1} e(t) \\
\frac{da(t)}{dt} &= t_{\text{latent}}^{-1} e(t) - \kappa(t) a(t) - \rho a(t)\\
\frac{di(t)}{dt} &= \kappa(t) a(t) - \beta i (t) - \mu i(t)  \\ 
\frac{dr(t)}{dt} &= \rho a(t)\ + \beta i (t) - \gamma r(t) \\ 
\frac{dp(t)}{dt} &= \mu i(t)  \label{eq:ode2}
\end{align} 
where $\alpha_a(t)$ and $\alpha_i(t)$ are the rates of exposure to the virus from the populations of asymptomatic (or unconfirmed) and confirmed infected subjects, respectively.
These two rates of exposure are defined as time-varying and independent model inputs to reflect different measures taken during the course of the pandemic.
Specifically, $\alpha_a(t)$ corresponds to exposure from asymptomatic carriers ($a$) and reflects social distancing and/or shelter-in-place strategies.
On the other hand, $\alpha_i(t)$ corresponds to exposure from infected subjects ($i$) and reflects quarantining of infected subjects.
We also model $\kappa(t)$, or the rate at which unconfirmed cases become confirmed, as a time-dependent input to reflect varying levels of screening and testing. 

The other parameters are assumed to be constant over the time horizons considered herein.
$t_\text{latent}^{-1}$ is the inverse of the latent period of the virus, or the time before an exposed subject becomes infectious. We assume a value of $t_\text{latent}^{-1} = 0.5\ \text{days}^{-1}$ based on Peng et al. \cite{peng2020}
$\rho$ describes the infectious period for subjects with unconfirmed infections, for which we assume a value of $\rho= 0.1\ \text{days}^{-1}$, based on Rockl\"{o}v et al. \cite{rocklov2020}
$\beta$ and $\mu$ describe the rates at which subjects with confirmed infections recover and perish, respectively.
Finally, $\gamma$ describes the rate at which recovered subjects become susceptible to the disease again. Since the immunity period for the virus is unknown, we assume a value of $\gamma = 0$.
While these parameters largely describe the virus itself, they may change over longer time horizons, e.g., as new treatments are developed.

In summary, the SEAIR model \eqref{eq:ode1}--\eqref{eq:ode2} has three time-varying inputs ($\alpha_a(t)$, $\alpha_i(t)$, and $\kappa(t)$), three parameters ($t_\text{latent}^{-1}$, $\rho$, and $\gamma$) whose values are based on available literature, and two parameters whose values must be estimated ($\beta$ and $\mu$).
Note that since $e(t)$ and $a(t)$ are unmeasured, their magnitudes relative to the infected population $i$ are effectively set by (the bounds on) $\kappa$.
As the prevalence of asymptomatic infection is established (e.g., Bendavid et al. \cite{Bendavid2020} estimate up to 4\% prevalence in Santa Clara County using serological testing) the magnitudes of $e$ and $a$ may be adjusted by scaling $\kappa$, and other parameters as relevant.

\subsection*{Parameter estimation results}

This work primarily addresses the current COVID-19 situation in the US, but the still relatively early stage of the outbreak in the US renders the available data insufficient to fit parameters reflective of prevention measures already in place. For this reason, we perform a comparative analysis by solving the parameter estimation problem for Italy, Spain, and Germany, where, while the epidemic has not been fully contained yet, the daily number of new confirmed cases has been on a steady decline.  

\subsubsection*{Model parameters for the US, Italy, Spain, and Germany}
\begin{figure}[!ht]
	\begin{center}
    \resizebox{0.85\textwidth}{!}{\input{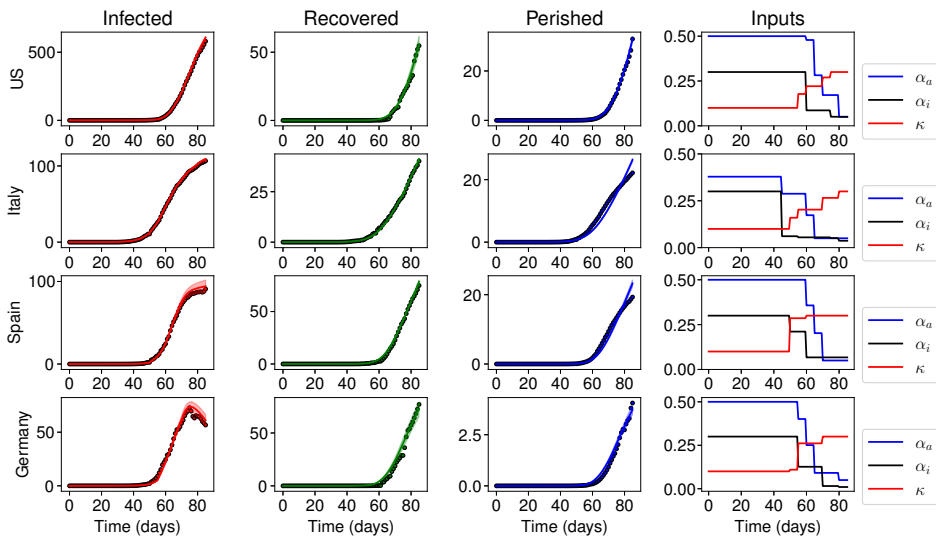}}
	\end{center}
	\caption{Historical data and model fit for infected $(i(t))$, recovered $(r(t))$, and perished $(p(t))$ subjects (in thousands). Solid lines represent the mean of 500 Monte Carlo simulations, shaded areas represent two standard deviations from the mean, and circle markers are historical data. The right-most column shows the fitted trajectories of the time-varying inputs.}
	\label{fig:parameter_fit}
\end{figure}

Figure \ref{fig:parameter_fit} shows the predicted values obtained by solving the parameter estimation problem and the historical data by country, retrieved by the Center for Systems Science and Engineering (CSSE) at Johns Hopkins University (\url{https://github.com/CSSEGISandData/COVID-19}; accessed April 16, 2020). Day 1 of the dataset corresponds to January 22, while day 85 corresponds to April 16. The solid line in each state trajectory plot shows the mean predicted value. Standard deviations for the predictions were estimated using bootstrapping, with the time-invariant parameters sampled from a multivariate normal distribution. The means and covariances of the estimated parameters for each country are provided in the Supplementary Information. 
Plots corresponding to infected, recovered and perished (left plots in Figure \ref{fig:parameter_fit}) were obtained by simulating \eqref{eq:ode1}--\eqref{eq:ode2} using the estimated trajectories of the time-varying inputs, shown in the right-most column of Figure \ref{fig:parameter_fit}.

Parameter estimation results provide several interesting insights.
First we note that the mortality rates, while higher for the European countries, are comparable in magnitude for the four countries considered and are similar to prior estimates. \cite{lai2020severe} The higher death rates in Italy and Spain are likely explained by demographic factors (e.g., age), as well as the medical resources available to treat the infected population.
However, the simulated trajectories of $p$ in Italy in Spain deviate slightly from the historical data, especially around day 80 (Fig \ref{fig:parameter_fit}), suggesting that the death rate $\mu$ may be time-varying.
While $\mu$ can vary in reality due to improved treatment, early detection bias, etc., it is difficult to anticipate future changes, and we therefore approximate it as a constant.

From the trajectories of $\alpha_a$, $\alpha_i$, and $\kappa$, we observe that social distancing measures (i.e., lower values of $\alpha_a$) were first implemented in Europe, and more recently the US, which agrees with their true chronology. 
This insight is confirmed by the fitted values for $e_0$, with Italy having the highest value and the US the lowest, which is representative of the beginning of the outbreak in each country.
In general terms, the identified evolution of containment and testing measures follow similar trends for all four countries, which supports that the model structure is correct and can appropriately reflect (see Methods) different dynamics of the same underlying disease. Further, we expect that the parameters obtained for the US, while being fitted using relatively premature information, are likely an adequate representation of the current COVID-19 situation. 

\subsubsection*{State simulations for fitted parameters}

\begin{figure}[!ht]
	\begin{center}
    \resizebox{0.85\textwidth}{!}{\input{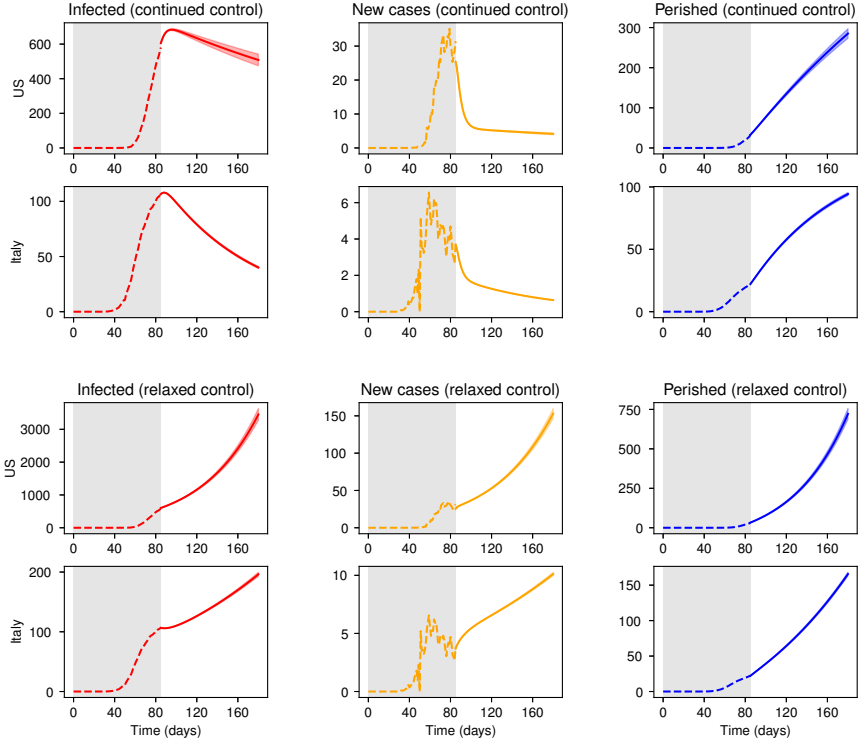}}
	\end{center}
	\vspace{-2mm}
	\caption{Simulation of future infected $(i)$, new confirmed cases $(\kappa a)$, and perished $(p)$ subjects (in thousands) for current control policies. Shaded grey area indicates historical data, color solid lines represent the mean of 500 Monte Carlo simulations, and color shaded area represents two standard deviations form the mean. }
	\label{fig:future_scenarios}
\end{figure}

For the estimated parameter values as described previously, we simulate the results of implementing two different simplistic control policies: (i) continuing with strict social distancing, quarantining, and testing, policies that result from continuing to lower the asymptomatic ($\alpha_a$) and infected ($\alpha_i$) exposures shown in Fig \ref{fig:parameter_fit}; and (ii) a relaxed policy with more lenient measures and reduced testing, in this cases the values of $\alpha_a$ and $\alpha_i$ are increased to 0.2 and 0.02, respectively,  while $\kappa$ is decreased to 0.2. The population levels resulting from implementing these two policies are shown  in Fig \ref{fig:future_scenarios}, where the numbers of recovered subjects are omitted for the sake of brevity. We additionally show the number of new confirmed cases per day, as it is a commonly used metric to illustrate the current spread/containment of the virus. Since the three European countries showed very similar trends under the two policies considered, we only compare the results for the US to those obtained for Italy. On the one hand, from Fig \ref{fig:future_scenarios} it is evident that relaxing current control policies can result in an alarming number of infected cases and deaths, particularly in the US. On the other hand, continuing with the strict shelter-in-place measures and maximizing testing can result in earlier and substantially flatter pandemic peaks, with significantly lower numbers of casualties. While the latter approach is the most effective at preventing further exponential spread of the disease, it is also the most socially and economically disruptive policy. 
In light of this trade-off, we argue that by means of employing optimal control concepts it is possible to find effective policies that maintain the number of infected cases below a given threshold, while minimizing the extent of social and economic disruption. 

\subsection*{Dynamic optimization results}
Here, we consider the optimal control of the COVID-19 infection in the US using the deterministic SEAIR model \eqref{eq:ode1}--\eqref{eq:ode2} with the mean values of the time-invariant parameters as described above. As such, the optimization problem only considers the mean predicted values (e.g., solid lines in Figs \ref{fig:parameter_fit} and \ref{fig:future_scenarios}). Future work may consider a stochastic optimization approach that, at the expense of increased computational effort, leverages information about model uncertainty.

\subsubsection*{Optimization of future actions}
We solve a min-max optimization problem that minimizes a measure of socioeconomic cost, subject to keeping the peak (max) value of the infected population below a given number $i_\text{peak}$. 
The lower bound of $\kappa(t)$ is raised to 0.15, from 0.10 in Figure \ref{fig:model}. This change was made because (i) the historical value of $\kappa$ for all countries analyzed quickly jumps to is maximum value 0.3, perhaps reflecting public initiatives, and (ii) the [0.15,0.3] range translates to $\frac{\rho}{\rho+\kappa} \in [25\%, 40\%]$, which is reflective of current estimates of the asymptomatic ratio. \cite{mizumoto2020, nishiura2020}

\begin{figure}[!ht]
	\begin{center}
    \resizebox{0.9\textwidth}{!}{\input{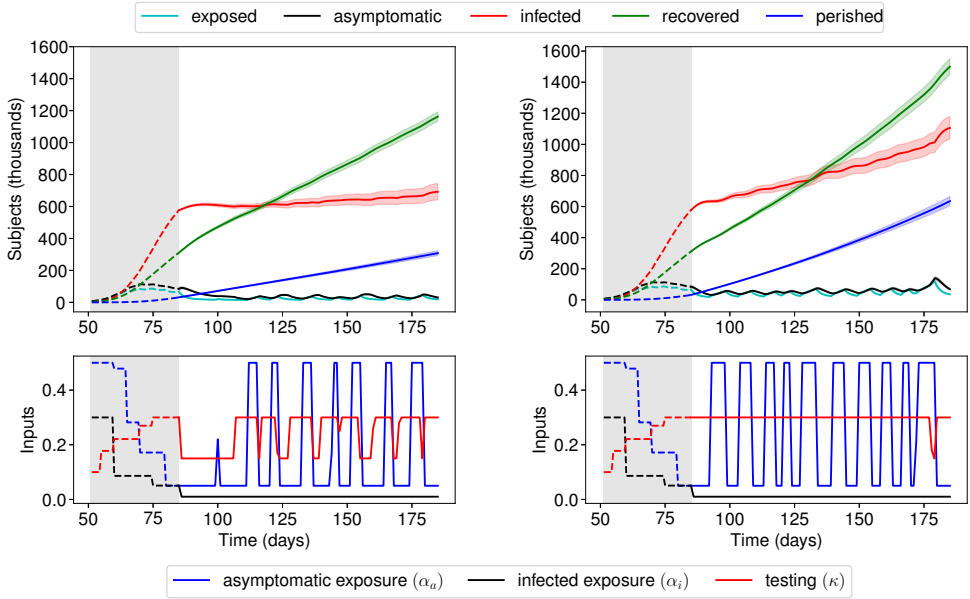}}
	\end{center}
	\caption{Optimal control policy to limit peak infections to 700,000 (left) or 1,400,000 (right) in the next 100 days. Top: population numbers, with two standard deviations shaded. Bottom: containment and testing profiles. The shaded grey area indicates past days, which were simulated using historical inputs (not optimized).}
    \label{fig:opti_5e5}
\end{figure}

Fig \ref{fig:opti_5e5} depicts the solution of this problem for the cases where $i_\text{peak}$ = 700,000 subjects and $i_\text{peak}$ = 1,400,000 subjects. The optimization problem was solved for the 100 days following the end of the available dataset ($t_0=85$, $t_f=185$), with the initial condition at $t_0$ computed by simulating days 1--85. Although the  optimization problem only considers the mean values (solid lines), two standard deviations are shown in the shaded areas to reflect parametric model uncertainty for the given inputs, computed using the same bootstrapping approach as above.

Considering the current status of the pandemic in the US, keeping the peak below 700,000 infected subjects is very challenging. Note that the minimum feasible value of $i_\text{peak}$ corresponds to a peak of 612,493 infected subjects.
Therefore, maintaining $i(t) \leq$ 700,000 requires immediately decreasing $\alpha_a$ and $\alpha_i$ to their lower bounds for approximately the next 25 days, during which the exposed population can decrease significantly.
After this period, $\alpha_a$ is relaxed to its upper bound for seven brief periods (Fig \ref{fig:opti_5e5}, left). However, $\alpha_i$ remains at its lower bound at all times, reflecting a strict quarantining policy.

This policy reveals that the impact of $\alpha_i$ on the infected population size is larger than that of $\alpha_a$.
In turn, this suggests that quarantining of infected people is more important than social distancing, which mitigates exposure to unconfirmed cases.
$\kappa$ is decreased during periods of decreased $\alpha_a$, suggesting that testing is less important during times of social distancing.
Intuitively, when the exposure to asymptomatic subjects is already low due to social distancing, there is less benefit to testing asymptomatic subjects (and transferring them to the similarly quarantined infected population).
On the other hand, $\kappa$ is increased preemptively for each period of increased $\alpha_a$, suggesting that testing is most important in the days leading up to a period of relaxed social distancing.
This testing moves asymptomatic subjects to the confirmed-infected population, which can remain quarantined.

Fig \ref{fig:opti_5e5} (right) depicts the solution obtained for a value of $i_\text{peak}$ = 1,400,000 people. 
While $\alpha_i$ again remains at its lower bound at all times, re-emphasizing the importance of quarantining infected subjects, there are more frequent and longer periods of increased $\alpha_a$.
In this case, $\kappa$ remains at its upper bound for most of the control horizon. 
In both cases, the optimization problem does not account for the effect(s) of the control policy after 100 days, and growth of the infected population near the end of the time horizon may be concerning.
This can be addressed, e.g., using a moving horizon control strategy, where policy measures are revised periodically, as discussed below.

\begin{figure}[!ht]
	\begin{center}
    \resizebox{0.5\textwidth}{!}{\input{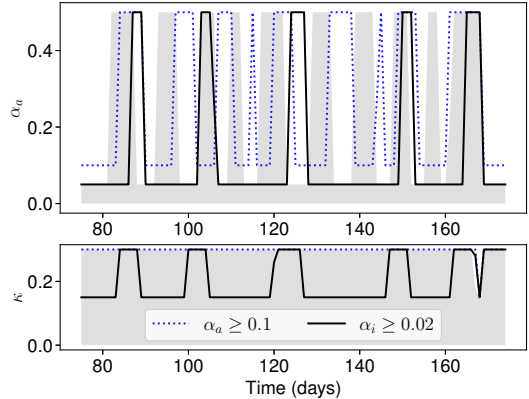}}
	\end{center}
	\caption{Optimal containment and testing strategies to limit peak infections to 1,000,000 in the next 100 days for different constraints on $\alpha_a$ and $\alpha_i$. Top: $\alpha_a(t)$. Bottom: $\kappa(t)$. The shaded grey area indicates the solution found using the normal bounds, replicated from Fig \ref{fig:opti_5e5} (right).}
	\label{fig:alpha_sensitivity}
\end{figure}

The lower bound used for $\alpha_i$ is lower than estimated values for the US for the past (Fig \ref{fig:opti_5e5}), corresponding to a new level of quarantining.
Furthermore, the current value of $\alpha_a$, or social distancing, may not be economically sustainable over longer periods.
To investigate the effect of not achieving these proposed levels, we solve the same problem with $i_\text{peak}$ = 1,400,000 people, using different lower bounds for the inputs.
Fig \ref{fig:alpha_sensitivity} shows the optimal input profiles found for the cases where the lower bound of $\alpha_a$ is doubled from 0.05 to 0.1 (dotted blue lines), and where the lower bound of $\alpha_i$ is doubled from 0.01 to 0.02 (solid black lines).
When the lower bound of $\alpha_a$ is doubled (less social distancing is achieved during lock-down periods), the same infected population peak can be maintained by increasing the frequency of social distancing periods.
However, when the lower bound of $\alpha_i$ is doubled (less quarantining is achieved), maintaining the same number of peak infections requires social distancing at almost all times, closely resembling the control policy found for a lower peak (Fig \ref{fig:opti_5e5}, left).
In either case, the optimal value of $\alpha_i$ remains at its lower bound (either 0.01 or 0.02) at all times and is therefore not shown.
Periods of social distancing are further increased in both frequency and duration if both bounds are doubled.

Compared to the decreases seen in Fig \ref{fig:future_scenarios} for the cases of continued control, the solutions in Fig \ref{fig:opti_5e5} seek to merely flatten the growth of the infected population.
The associated optimal control policies resemble ``bang-bang'' control (the inputs are always at the lower or upper bound), which can be expected from a systems-theoretic point of view for \eqref{eq:ode1}--\eqref{eq:ode2}, as the equations are linear functions of the inputs $\alpha_a$, $\alpha_i$, and $\kappa$.
This result supports the strategy proposed by the Imperial College COVID-19 Response Team, \cite{ferguson2020} which involves periodic suppression measures.
Their strategy triggers the start of a social distancing period by when the number of weekly ICU cases increases past an ``on'' threshold, and the end of the period when the same number decreases below an ``off'' threshold.
Fig \ref{fig:opti_5e5} shows that this type of strategy maintains a low exposed population, and therefore flattens the growth of the infected population while still allowing periods of social mobility.

\subsubsection*{Revisiting past actions}

\begin{figure}[!ht]
	\begin{center}
    \resizebox{0.9\textwidth}{!}{\input{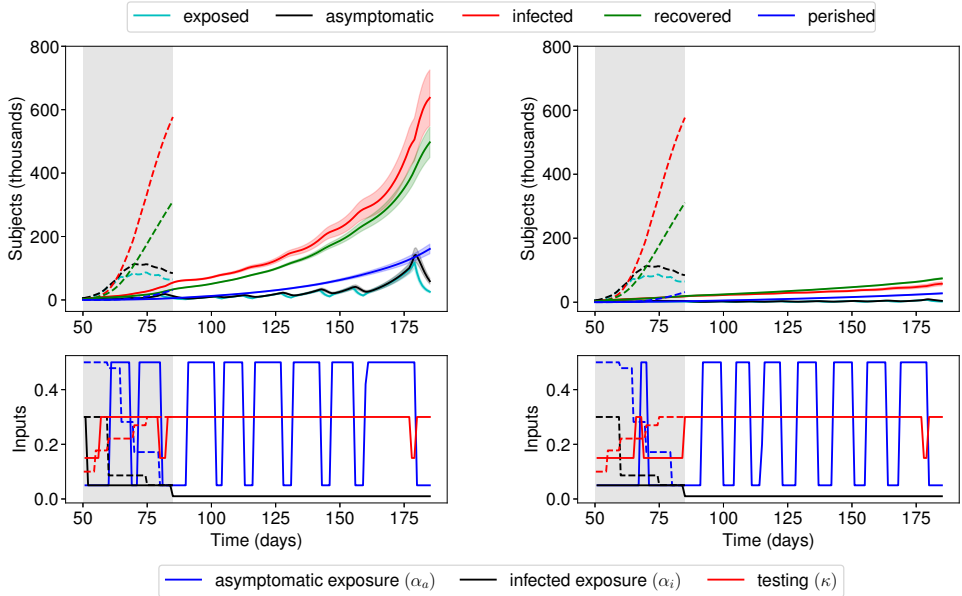}}
	\end{center}
	\caption{Optimal control policy to limit peak infections to 700,000 (left) or 70,000 (right) in the past 35 days and next 100 days. Top: population numbers, with two standard deviations shaded. Bottom: containment and testing profiles. The shaded grey area indicates past days, for which the true historical inputs and outputs are shown as dashed lines and the optimized are shown as solid lines.}
	\label{fig:opti_5e5_past}
\end{figure}

We examine the same optimal control problem with a time horizon starting from day 50 (i.e., covering 35 days in the past).
Rolling the time horizon of the optimization problem backwards to $t_0=50$, allows us to investigate the optimal inputs for days 50--85, corresponding roughly to the latter half of March and the first half of April.
Fig \ref{fig:opti_5e5_past} (left) depicts the solution obtained for a value of $i_\text{peak}$ = 700,000 people. The initial condition at $t_0=50$ was computed by simulating days 1--50.
The minimum historical values for $\alpha_a$ and $\alpha_i$ were used as their respective lower bounds for days 50--85, such that the optimal control policy for days 50--85 only involves an extent of quarantining and social distancing already experienced by the general population.

In this case, maintaining the number of infected subjects below 700,000 appears easy.
The minimum feasible peak of infected subjects is 16,543, compared to 612,493 when the optimization starts at day 85.
Therefore, maintaining $i(t) \leq 700,000$ can involve extended periods of no social distancing $\alpha_a=0.5$.
Similar to the result in Fig \ref{fig:opti_5e5} (right), testing and quarantine are important in this scenario: $\kappa$ remains at its upper bound and $\alpha_i$ its lower bound.
The solution for an order-of-magnitude lower minimum peak, corresponding to 70,000 infected subjects is shown in Fig \ref{fig:opti_5e5_past} (right). 
The optimal control policy for this scenario involves more frequent periods of social distancing.
Early implementation of testing and quarantining strategies clearly has an enormous effect, manifest in the large difference between the historical inputs (dashed) and optimal inputs (solid) in days 50--85 of Fig \ref{fig:opti_5e5_past}.

\subsubsection*{Estimation of hidden states and moving horizon control}

\begin{figure}[!ht]
	\begin{center}
    \resizebox{0.9\textwidth}{!}{\input{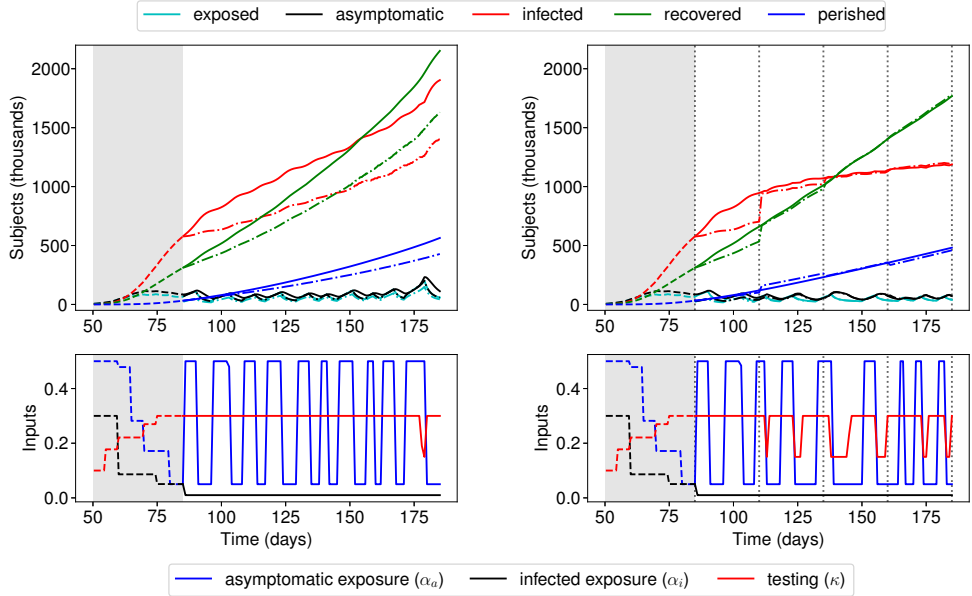}}
	\end{center}
	\caption{Optimal moving horizon control policy (right) to limit peak infections to 1,400,000, with $e$ and $a$ underestimated by a factor of three at $t=74$, and comparison to the same situation without a moving horizon strategy (left). Top: predicted (dash-dotted) and true (solid) population numbers. Bottom: containment and testing profiles. The shaded grey area indicates past days, which were simulated using historical inputs (not optimized). The policies are updated every 25 days with daily state estimation.}
	\label{fig:MH}
\end{figure}

The SEAIR model \eqref{eq:ode1}--\eqref{eq:ode2} includes two \textit{hidden} states, $e(t)$ and $a(t)$, which are not measured in practice (note that widespread serological testing\cite{Bendavid2020} may eventually reveal the true levels of the asymptomatic population $a$).
Nevertheless, solutions for the optimal control problem based on \eqref{eq:ode1}--\eqref{eq:ode2} are highly dependent on the values of $e(t)$ and $a(t)$, and therefore the initial conditions $e(t_0)$ and $a(t_0)$.
Fig \ref{fig:MH} (left) depicts the optimal control policies for $i_\text{peak} = 1,400,000\ \text{people}$ over the next 100 days (similar to Fig \ref{fig:opti_5e5}, right), where $e(t_0)$ and $a(t_0)$ are now underestimated by a factor of three. 
This is a practically motivated scenario, as the number of asymptomatic cases is considered largely uncertain.~\cite{gostic2020estimated, anderson2020will}
The dashed-dotted lines show the state profiles predicted by the optimization problem (with incorrect initial conditions), while the solid lines show their true evolution for the given inputs.

The error in the hidden states $e(t)$ and $a(t)$ causes the predicted and actual profiles to diverge over time.
By the end of the 100 days, there are 1.90 million actual infected subjects, almost 40\% greater than the predicted 1.40 million. 
To account for discrepancies between the modeled and true systems, the optimal control inputs should be periodically updated.
We propose to achieve this with a \textit{moving horizon} decision-making strategy: after a given length of time, the dynamic optimization problem is re-solved, setting the measured values of the states as their initial conditions.
The values of the un-measured, hidden states can be estimated using state estimation techniques, such as the Kalman filter (the optimal linear estimator) or its nonlinear extensions.

We consider a relatively long time span of 25 days before each revision of policy decisions (and re-optimization), as planned suppression strategies likely cannot be altered quickly.
Fig \ref{fig:MH} (right) shows the optimal trajectories found for the same problem (initial conditions for $e$ and $a$ underestimated by a factor of three) using the moving horizon strategy.
The vertical dotted lines indicate times when the optimization problem was resolved.
Values of the hidden states were estimated daily with the (discrete) unscented Kalman filter, \cite{wan2000} implemented using the pykalman library (\url{https://pykalman.github.io/}; Accessed April 7, 2020).
Measurements for state estimation and control updates were simulated by adding independent, normally distributed noise to the true state values.
A standard deviation of 5,000 people was used for the measured states ($i$, $r$, and $p$) based on the size of residuals during parameter estimation.

The predicted values (dash-dotted lines in Fig \ref{fig:MH}) are first updated when the optimization problem is re-solved at $t$=110 days.
The solutions to the re-optimization problems are shown in the Supplementary Information.
The effect of this moving horizon strategy is quite significant, in comparison to the optimal policy shown in Fig \ref{fig:MH} (left).
Periods of social distancing are quickly increased after day 110 to account for the larger-than-expected increase in infected population.
$\kappa$ is also updated accordingly.
The moving horizon approach exhibits three clear benefits:
\begin{itemize}
    \item Errors between the model predictions and reality are minimized by incorporating new measurements as they become available
    \item The infected population does not grow exponentially at the end of the time horizon, as is the case when considering only a single time horizon
    \item The control policies can be computed indefinitely into the future by shifting the time window forward in time, while the problem size remains tractable (only 100 days are considered during optimization)
\end{itemize}

The effect of state estimation is most notable in the first moving horizon window, between days 85--110.
Here, while the values predicted by the optimization problem are consistently under-predicted, the estimated values approach the (unknown) true values as daily measurements are added.
The estimated values over time are shown in the Supplementary Information.
The new solution of the optimization problem at day 110 uses the population levels measured/estimated based on new data as initial conditions, and therefore predicts more accurate trajectories.

\section*{Discussion}
This work investigated dynamic optimization strategies to characterize and control the US COVID-19 outbreak, by minimizing the socioeconomic cost associated with containment strategies and testing.
The results provide several overarching conclusions.

The quarantining of infected subjects is the most important of the considered mitigation strategies and should be maximized at all times. 
Additionally, periods of social distancing help to flatten the peak by preventing exposure from asymptomatic and unconfirmed cases.
Screening and testing for the disease are key immediately preceding periods of relaxed social distancing, in order to minimize the number of unconfirmed infections during periods of social mobility.
Early action has much larger effects than later interventions, even as the later interventions are more drastic.
Optimal policies are highly dependent on estimates of ``hidden states,'' i.e., the asymptomatic and unconfirmed cases.
Moving horizon (periodically updated) policies and state estimation should be used to mitigate inaccuracies in the model and counts of asymptomatic/unconfirmed cases, by accounting for new data as they becomes available. 

The ``on-off'' policies identified as optimal are characteristic of the class of problems considered (linear objective function and input-affine nonlinear model).
Further, these policies are likely the easiest to implement in practical scenarios and to convey to the general population (as opposed to a policy where the social distancing parameters would take values between their upper and lower bounds). Their implementation would simply alternate between the strictest possible limitations, followed by periods of relative freedom of movement. 

The model used in this work does not include population influx to and/or outflux from the given system.
Therefore, the optimal containment and testing strategies found do not account for new cases that may enter from outside the US.
We also assumed that the recovery and death rates for the virus are constant and equal to their historical values for the next 100 days.
Thus the results do not account for the possibility of improved medical treatments, vaccine development, viral mutation, etc.
Similarly, the model does not account for possible surveillance/detection bias in the historical data (e.g., increased likelihood of testing for subjects with more severe symptoms).

\section*{Methods}
\subsection*{Parameter estimation}

We follow nonlinear (least-squares) regression for parameter estimation, which can be cast as an optimization problem, where the objective is to find the parameter values that minimize the mean squared error (MSE) between the predicted states and their measured values.
The measured values correspond to data retrieved by the Center for Systems Science and Engineering (CSSE) at Johns Hopkins University (\url{https://github.com/CSSEGISandData/COVID-19}; accessed April 16, 2020) and are provided in the Supplementary Information for several countries. The data consist of the total number of infected ($\hat{i}_j$), recovered ($\hat{r}_j$) and dead ($\hat{p}_j$) reported subjects, where $j$ represents each day during the time period from January 22 to April 16, where the index $j$ represents the time period each value was recorded. The MSE is given by: 
\begin{equation}
L_0 = \sum_{j=0}^N (i(t=j) - \hat{i}_j )^2 + (r(t=j) - \hat{r}_j )^2 + (p(t=j) - \hat{p}_j )^2 
\label{eq:pe_loss_fn}
\end{equation}
We note that when solving the parameter estimation problem, the contribution of the term  $\rho a(t)$ to the recovered state $r(t)$, is removed from the system \eqref{eq:ode1}-\eqref{eq:ode2}. This is because the data obtained only accounts for those recovered that were confirmed to have the disease, meaning the only contribution to the observed recovered is the term  $\beta i(t)$.

The parameters with values/trajectories to estimate are $\alpha_a(t)$, $\alpha_i(t)$, $\kappa(t)$, $\beta$, and  $\mu$. Furthermore, we must estimate the initial condition of the unreported states $e$ and $a$ at $t=0$. We assume that there are initially no asymptomatic infections. The ``seed'' of the outbreak is therefore the initial number of exposed subjects $e_0 = e(t=0)$, which we include as a parameter in the least-squares problem. 
We include bounds on the possible parameter values, based on values reported for similar models fitted to data from other regions. \cite{peng2020}
Note that, while $\kappa$ reflects the level of testing, it also affects the predicted asymptomatic ratio, which cannot be controlled.
Therefore, tighter bounds should be used for $\kappa$ than for $\alpha_a$ and $\alpha_i$.
The estimated parameters and their bounds are summarized in Figure\ref{fig:model}.

To prevent over-fitting the available data, we restrict the time-varying inputs ($\alpha_a$, $\alpha_i$, and $\kappa$) to be piece-wise constant over five-day intervals. 
These constraints reflect the fact that policies implemented have significant time delays in steering the states towards the desired values, and therefore should not and practically cannot be manipulated too frequently before their effect is observed on the population.
Additionally, we constrain $\alpha_a(t)$ and $\alpha_i(t)$ to be monotonically decreasing over time, reflecting the enforcement of increasingly stricter disease control measures.  Similarly, $\kappa(t)$ was constrained to be monotonically increasing over time to account for increased COVID-19 screening availability. 

To easily interface with raw data sources, the least-squares regression problem was implemented in Python using the Pyomo modeling and optimization package. \cite{hart2017pyomo} We discretized the dynamical system  \eqref{eq:ode1}-\eqref{eq:ode2} with respect to the time domain using orthogonal collocation on finite elements, \cite{biegler2007overview} with one finite element per day. We used IPOPT \cite{ipopt} to solve the resulting nonlinear dynamic optimization problem to local optimality. 
Standard deviations for the time-invariant parameters were approximated by fixing the time-varying inputs to their estimated trajectories, and solving a maximum likelihood problem for the time-invariant parameters using the Model Validation tool in gPROMS (general PROcess Modeling System) v5.1.4. \cite{gproms} Residuals were assumed independent and normally distributed, with variances estimated in the same maximum likelihood problem. The parameter covariances (see SI) are then computed from the diagonal entries of the Hessian of the objective function.

\subsection*{Optimal control}
The aim of the optimal control problem is to find the trajectories of ``handles'' $\alpha_a(t)$,  $\alpha_i(t)$ and $\kappa(t)$, that minimize (or maximize) the value of a certain objective function.
Most response measures for the COVID-19 outbreak seek to ``flatten'' the epidemic curve, that is, to contain the growth rate of the number of infected subjects via a combination of social distancing and testing. Clearly, social distancing and isolation/quarantining carry significant social and economic costs. 
We thus formulate the optimal control problem as a dynamic optimization problem, aiming to minimize a measure of social and economic cost subject to ensuring that the maximum number of infected subjects remains under under a given peak value, $i_\text{peak}$.
The optimization problem is expressed mathematically as:

\begin{equation}
    \begin{aligned}
    \underset{\alpha_a(t), \alpha_i(t), \kappa(t)}{\text{min}} &\int_{t_0}^{t_f} C(t) dt \\
    \text{s.t.} & \quad \text{SEAIR model \eqref{eq:ode1}-\eqref{eq:ode2}} \\
     & \quad \underset{t}{\text{max}}(i(t)) \leq i_\text{peak} \\
     & \quad C(t) = - \alpha_a(t) - \alpha_i(t) + \lambda_\kappa \kappa(t) \\
     & \quad \alpha_a(t) \in [0.05, 0.5] \\
     & \quad \alpha_i(t) \in [0.01, 0.3] \\
     & \quad \kappa(t) \in [0.15, 0.3]
    \end{aligned}
    \label{eq:opt_problem}
\end{equation}
where $C(t)$ is the cost function, and $\lambda_\kappa$ is the relative cost of testing (increasing $\kappa$). Testing costs are assumed to be relatively small in comparison to isolation measures, and we selected $\lambda_\kappa = 0.1$. 
The solution to (\ref{eq:opt_problem}) can provide control policies over the full time horizon $[t_0,t_f]$ as in the case of the initial studies presented in this work. 
Alternatively, the policies can be updated after a shorter horizon (before $t_f$ is reached), as in the case of the moving horizon approach. In the latter case, the time window considered, $[t_0 , t_f]$ is ``shifted right'' at a pre-determined time interval. Note that $t_f$ is typically longer than the frequency at which the optimization problem is recomputed, and only the solution for the first time step(s) is implemented.
For example, the moving horizon scenario presented here is solved with $t_f$ = 100 days, but with policies updated after each 25-day period.


While the problem \eqref{eq:opt_problem} identifies the minimum societal cost required to achieve a certain peak value of infections, a similar problem could be formulated by minimizing $i_\text{peak}$ subject to an upper bound on $C(t)$, i.e., finding the lowest achievable peak for a given total societal cost.
For this study we investigate the relationship between $C$ and $i_\text{peak}$ by solving \eqref{eq:opt_problem} for varying values of $i_\text{peak}$, provided in the Supplementary Information.

The optimization problem \eqref{eq:opt_problem} was again solved by discretization of the time domain using orthogonal collocation, with one finite element per day. To report the best possible solutions, the resulting algebraic optimization problem was implemented in GAMS and solved using the commercial global optimization solver BARON v18.5.8. \cite{tawarmalani2005}
Problems were solved to a 0.1\% optimality gap (i.e., the reported solution is proven to have an objective function value within 0.1\% of that of the best possible solution).
We additionally provide an implementation of the same problem using open-source tools (Pyomo \cite{hart2017pyomo} and IPOPT \cite{ipopt}), which can be used to solve \eqref{eq:opt_problem} to local optimality.

\section*{Additional information}

\textbf{Data Availability} Models are available at \url{https://github.com/Baldea-Group/covid-19}. \\

\noindent
\textbf{Competing interests} The authors declare no competing interests. \\

\bibliographystyle{unsrt}
\bibliography{covidrefs}

\section*{Acknowledgements}

Partial financial support from the National Science Foundation (NSF) through CAREER Award 1454433 is acknowledged with gratitude.
The authors also thank Dr Richard Pattison (Apeel Sciences) for his insightful comments throughout this study.

\section*{Author contributions statement}

All authors conceived the experiments; C.T. and F.L. implemented the optimisation models, conducted the computational experiments, and analysed the results; All authors prepared and reviewed the manuscript. 

\newpage
\section*{Supplementary Information}

\renewcommand{\thefigure}{S\arabic{figure}}
\renewcommand{\thetable}{S\arabic{table}}
\setcounter{figure}{0}  

A least-squares regression problem was solved for the values of the parameters in the SEAIR model (1)--(6). The estimated means of the parameters for the USA, Italy, Spain, and Germany are given below in Table \ref{table_tiv_params}, and the covariance matrices are given in Table \ref{tiv_covariances}.
These covariance matrices were used to generate samples for bootstrapping.

\begin{table}[!ht]
	\centering
	\begin{tabular}{|l|l|l|l|l|}
		\hline
		& {\bf USA} & {\bf Italy} & {\bf Spain} & {\bf Germany} \\ \hline
		$\mu$ [days$^{-1}$] & $ 4.405 \times 10^{-3}$ & $10.619 \times 10^{-3}$ & $ 11.871 \times 10^{-3}$ & $2.402 \times 10^{-3}$ \\ \hline
		$\beta$ [days$^{-1}$] & $7.467 \times 10^{-3}$ & $16.644 \times 10^{-3}$ & $40.129 \pm \times 10^{-3}$ & $46.838 \times 10^{-3}$\\ \hline
		$e_0$ [people] & $1.684 \times 10^{-1}$ & $90.035 \times 10^{-1}$ & $2.004 \times 10^{-1}$ & $1.751 \times 10^{-1}$\\ \hline
	\end{tabular}
	\begin{flushleft} 
	\end{flushleft}
	\caption{Mean values of time-invariant model parameters.}
	\label{table_tiv_params}
\end{table}

\begin{table}[!ht]
	\centering
	\begin{tabular}{|l|l|l|l|} \hline
		\multicolumn{4}{|l|}{\bf USA} \\ \hline
		$\beta$ [days$^{-1}$] & 1.39 $\times 10^{-7}$ & & \\ \hline
		$e_0$ [people] & 4.54 $\times 10^{-7}$ & 2.61 $\times 10^{-6}$ & \\ \hline
		$\mu$ [days$^{-1}$] & -3.59 $\times 10^{-9}$ & -2.75 $\times 10^{-8}$ & 8.59 $\times 10^{-9}$\\ \hline
	\end{tabular}
	
	\vspace{10pt}
	\begin{tabular}{|l|l|l|l|} \hline
		\multicolumn{4}{|l|}{\bf Italy} \\ \hline
		$\beta$ [days$^{-1}$] & 3.00 $\times 10^{-9}$ & & \\ \hline
		$e_0$ [people] & -2.55 $\times 10^{-8}$ & 8.09 $\times 10^{-4}$ & \\ \hline
		$\mu$ [days$^{-1}$] & -2.58 $\times 10^{-9}$ & 3.43 $\times 10^{-6}$ & 1.79 $\times 10^{-8}$\\ \hline
	\end{tabular}
	
	\vspace{10pt}
	\begin{tabular}{|l|l|l|l|} \hline
		\multicolumn{4}{|l|}{\bf Spain} \\ \hline
		$\beta$ [days$^{-1}$] & 4.44 $\times 10^{-8}$ & & \\ \hline
		$e_0$ [people] & 1.12 $\times 10^{-7}$ & 8.08 $\times 10^{-7}$ & \\ \hline
		$\mu$ [days$^{-1}$]  & -8.89 $\times 10^{-9}$ & 9.17 $\times 10^{-8}$ & 3.43 $\times 10^{-8}$\\ \hline
	\end{tabular}
	
	\vspace{10pt}
	\begin{tabular}{|l|l|l|l|} \hline
		\multicolumn{4}{|l|}{\bf Germany}\\ \hline
		$\beta$ [days$^{-1}$] & 6.36 $\times 10^{-6}$ & & \\ \hline
		$e_0$ [people]& 1.47 $\times 10^{-5}$ & 5.49 $\times 10^{-5}$ & \\ \hline
		$\mu$ [days$^{-1}$] & 5.47 $\times 10^{-8}$ & -1.45 $\times 10^{-7}$ & 4.66 $\times 10^{-9}$\\ \hline
	\end{tabular}
	
	\caption{Covariance matrices of the time-invariant model parameters.}
	\label{tiv_covariances}
\end{table}

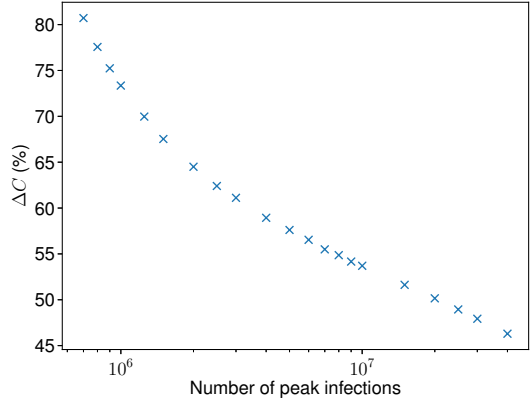
\begin{figure}[!ht]
	\centering
	\raisebox{-0.6\height}{\resizebox{0.5\textwidth}{!}{\input{costfunction.pgf}}}
	\begin{tabular}{|l|l|} \hline
		$i_\text{peak}$ & $\Delta C$ (\%) \\ \hline
		4 $\times 10^7$ & 46.29 \\
		3 $\times 10^7$ & 47.92 \\
		2.5 $\times 10^7$ & 48.94 \\
		2 $\times 10^7$ & 50.16 \\
		1.5 $\times 10^7$ & 51.62 \\
		1 $\times 10^7$ & 53.70 \\
		9 $\times 10^6$ & 54.16 \\
		8 $\times 10^6$ & 54.87 \\
		7 $\times 10^6$ & 55.50 \\
		6 $\times 10^6$ & 56.53 \\
		5 $\times 10^6$ & 57.61 \\
		4 $\times 10^6$ & 58.93 \\
		3 $\times 10^6$ & 61.11 \\
		2.5 $\times 10^6$ & 62.41 \\
		2 $\times 10^6$ & 64.50 \\
		1.5 $\times 10^6$ & 67.53 \\
		1.25 $\times 10^6$ & 69.97 \\
		1 $\times 10^6$ & 73.35 \\
		9 $\times 10^5$ & 75.24 \\
		8 $\times 10^5$ & 77.57\\
		7 $\times 10^5$ & 80.71 \\ \hline
	\end{tabular}
	\caption{Cost required to limit peak infections to various values. Costs given as percentage increases from no-action-taken scenario.}
	\label{fig:costfunction}
\end{figure}

Given the parameters for the US outbreak, the optimization problem (8) can be solved for the next 100 days.
Fig 4 shows the solutions for peak infected population $i_\text{peak}$ values of 700,000 and 1,400,000 people.
Here, we solve (8) for various values of $i_\text{peak}$ to investigate the relationship between $i_\text{peak}$ and the associated socioeconomic cost $C$.
The results are given in Fig \ref{fig:costfunction}.
As the cost function is in arbitrary units, we report the cost here as the percentage increase over the cost function value for normal societal operation, i.e., with no mitigation/testing strategies ($\alpha_a=0.5$, $\alpha_i=0.3$, $\kappa=0.1$).

\begin{figure}[!ht]
	\begin{center}
		\resizebox{0.6\textwidth}{!}{\input{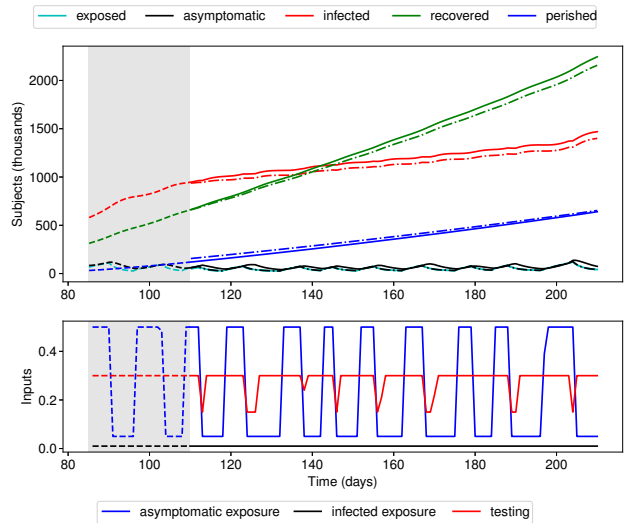}}
	\end{center}
	\caption{Optimal control policy found at day 110 in moving horizon scenario. Top: predicted (dash-dotted) and true (solid) population numbers. Bottom: containment and testing profiles. The shaded grey area indicates past days, which were simulated using historical inputs (not optimized).}
	\label{fig:MH1}
\end{figure}

The moving horizon strategy shown in Fig 6 involves updating control policies in 25-day intervals (after day 85).
The problem solved at each update is identical, except that the initial conditions are updated based on the current measured/estimated populations.
The solutions obtained at days 110, 135, and 160 are shown, respectively, in Fig \ref{fig:MH1}, Fig \ref{fig:MH2}, and Fig \ref{fig:MH3}.

\begin{figure}[!ht]
	\begin{center}
		\resizebox{0.6\textwidth}{!}{\input{US_opt_MH2.pgf}}
	\end{center}
	\caption{Optimal control policy found at day 135 in moving horizon scenario. Top: predicted (dash-dotted) and true (solid) population numbers. Bottom: containment and testing profiles. The shaded grey area indicates past days, which were simulated using historical inputs (not optimized).}
	\label{fig:MH2}
\end{figure}

\begin{figure}[!ht]
	\begin{center}
		\resizebox{0.6\textwidth}{!}{\input{US_opt_MH3.pgf}}
	\end{center}
	\caption{Optimal control policy found at day 160 in moving horizon scenario. Top: predicted (dash-dotted) and true (solid) population numbers. Bottom: containment and testing profiles. The shaded grey area indicates past days, which were simulated using historical inputs (not optimized).}
	\label{fig:MH3}
\end{figure}

\begin{figure}[!ht]
	\begin{center}
		\resizebox{0.6\textwidth}{!}{\input{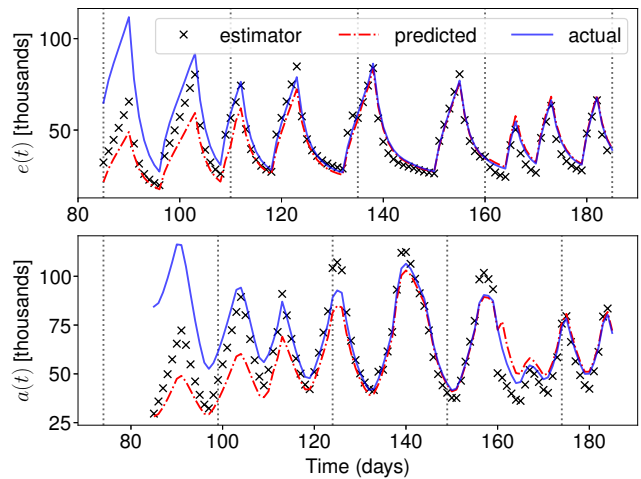}}
	\end{center}
	\caption{Estimation of hidden states during moving horizon control. Top: estimated, predicted (during optimization), and actual values of $e(t)$. Bottom: estimated, predicted, and actual values of $a(t)$. Re-optimization is performed every 25 days, while state estimation is performed daily.}
	\label{fig:estimator}
\end{figure}

The moving horizon strategy (Fig 6) relies on state estimation for values of $e(t)$ and $a(t)$. While the control policies are only updated every 25 days, the state estimates are updated daily as new measurements are made available.
The values of $e(t)$ and $a(t)$ were estimate using an unscented Kalman filter for moving horizon control.
The estimates over time are shown in Fig \ref{fig:estimator}.

\end{document}